\documentclass[pre,amsmath,amssymb,aps,twocolumn]{revtex4-1}

\usepackage{color}

\usepackage{graphicx}% Include figure files
\usepackage{dcolumn}% Align table columns on decimal point
\usepackage{bm}% bold math
\usepackage{epstopdf}
\usepackage{color}
\usepackage[colorlinks = true, citecolor = blue, linkcolor = blue, urlcolor=blue]{hyperref}

\begin{document}

\preprint{APS/123-QED}

\title{Phase transitions in electron spin resonance under continuous microwave driving
}

\author{A. Karabanov, D.C. Rose, W. K\"ockenberger, J.P. Garrahan, and I. Lesanovsky}

\affiliation{School of Physics and Astronomy, University of Nottingham, \\University Park, NG7 2RD, Nottingham, UK}
\affiliation{Centre for the Mathematics and Theoretical Physics of Quantum Non-equilibrium Systems,
University of Nottingham, Nottingham NG7 2RD, UK}

\date{\today}% It is always \today, today,
             %  but any date may be explicitly specified

\begin{abstract}
We study an ensemble of strongly coupled electrons under continuous microwave irradiation interacting with a dissipative environment, a problem of relevance to the creation of highly polarized non-equilibrium states
in nuclear magnetic resonance. We analyse the stationary states of the dynamics, described within a Lindblad master equation framework, at the mean-field approximation level. This approach allows us to identify steady state phase transitions between phases of high and low polarization controlled by the distribution of disordered electronic interactions. We compare the mean-field predictions to numerically exact simulations of small systems and find good agreement. Our study highlights the possibility of observing collective phenomena, such as metastable states, phase transitions and critical behaviour in appropriately designed paramagnetic systems. These phenomena occur in a low-temperature regime which is not theoretically tractable by conventional methods, e.g., the spin-temperature approach.
\end{abstract}

\maketitle

\noindent
{\bf \em Introduction} --- The control and detection of magnetization arising from a polarized ensemble of unpaired electron spins forms the basis of electron spin, or paramagnetic, resonance  (ESR/EPR); a powerful spectroscopy tool for studying paramagnetic materials placed in a static external magnetic field. The underpinning key principle for this technique is the application of oscillating magnetic fields close to or at the electronic Larmor frequency (usually in the microwave regime) to generate non-equilibrium distributions of populations and coherences between quantum states that lead to detectable signals \cite{z-45,l-67,sj-01}.  The evolution of systems of isolated or only weakly coupled paramagnetic centres under the effect of these fields is well understood.  A more challenging problem is to predict the response of strongly coupled electron ensembles to such perturbations, particularly in samples in the solid state in which anisotropic components of the electronic interactions are not averaged out by thermal motion. Insight into the dynamics of strongly coupled, microwave driven electronic ensembles is also needed in order to improve our understanding of dynamic nuclear polarization (DNP), which is an out-of-equilibrium technique to enhance the sensitivity of nuclear magnetic resonance (NMR) applications by orders of magnitude (see, e.g., Ref. \cite{gps,ak-12,w-16}) --- in particular, this concerns the cross effect and thermal mixing DNP mechanisms \cite{klm-63,h-67,g-04,b-68,ar-72,ag-82,k-16}.

Here we shed light on the non-equilibrium stationary states of a strongly interacting electronic ensemble under continuous microwave driving and subject to dissipation to the environment. We model the dynamics of this system in terms of a Markovian master equation and use a mean-field approximation to compute the steady state phase diagram. This reveals phase transitions between states of high and low electronic polarisation as well as the emergence of a critical point that displays Ising universality \cite{m}. These features are controlled by the distribution of the disordered electronic spin-spin interactions. The uncovered mean-field transitions imply the emergence of  metastable states and accompanying intermittent dynamics \cite{Ates2012Ising,Rose2016,foss2016emergent}, which we confirm numerically through simulations of small systems. Our results suggest that under appropriate conditions collective phenomena such as metastability, phase transitions and critical behaviour should be observable in driven-dissipative, paramagnetic systems. These predictions complement those of conventional theoretical approaches, based, e.g., on the so-called \emph{spin-temperature} which, due to their restriction to certain parameter regimes, would only predict a homogenous quasi-equilibrium state  \cite{p-62,b-68,ar-72,a-78,ag-82,j-12,v-13,sr-12,lr-15}.

\noindent
{\bf \em Model} --- We model the evolution of the electron system within the framework of a Markovian Lindblad master equation. The density matrix $\rho$ of a system consisting of $N$ microwave-driven electrons evolves according to $\dot\rho=-i[H,\rho]+\mathcal{D}\rho$. The Hamiltonian $H$ at high static magnetic field, in the rotating frame approximation, is given by
\begin{eqnarray}
H&=&\sum_k\left(\omega_1S_{kx}+\Delta_kS_{kz}\right)+3\sum_{k<k'}D_{kk'}S_{kz}S_{k'z} \nonumber
\\
&&-\sum_{k<k'}D'_{kk'}{\mathbf S}_k\cdot{\mathbf S}_{k'}.
\label{h}
\end{eqnarray}
Here $\omega_1$ is the strength of the microwave field, $\Delta_k$ are the offsets of the electron Larmor frequencies (detunings) from the microwave carrier frequency, and $D_{kk'}$, $D'_{kk'}$ are coefficients that parametrize the strength of the anisotropic and isotropic parts of the spin-spin dipolar and exchange interactions \cite{sj-01}. Depending on the degree of order and symmetries within the sample structure, $D_{kk'}$ and $D'_{kk'}$ can either be well defined (e.g., for crystals) or considered to be random (e.g., for glasses). In amorphous materials $\Delta_k$ are also distributed
due to the anisotropic interaction of the electrons with the static field, leading to inhomogeneous broadening of the EPR line \cite{sj-01,pf-79,k-16}.

\begin{figure*}[t]
\begin{center}
\includegraphics[scale=.3]{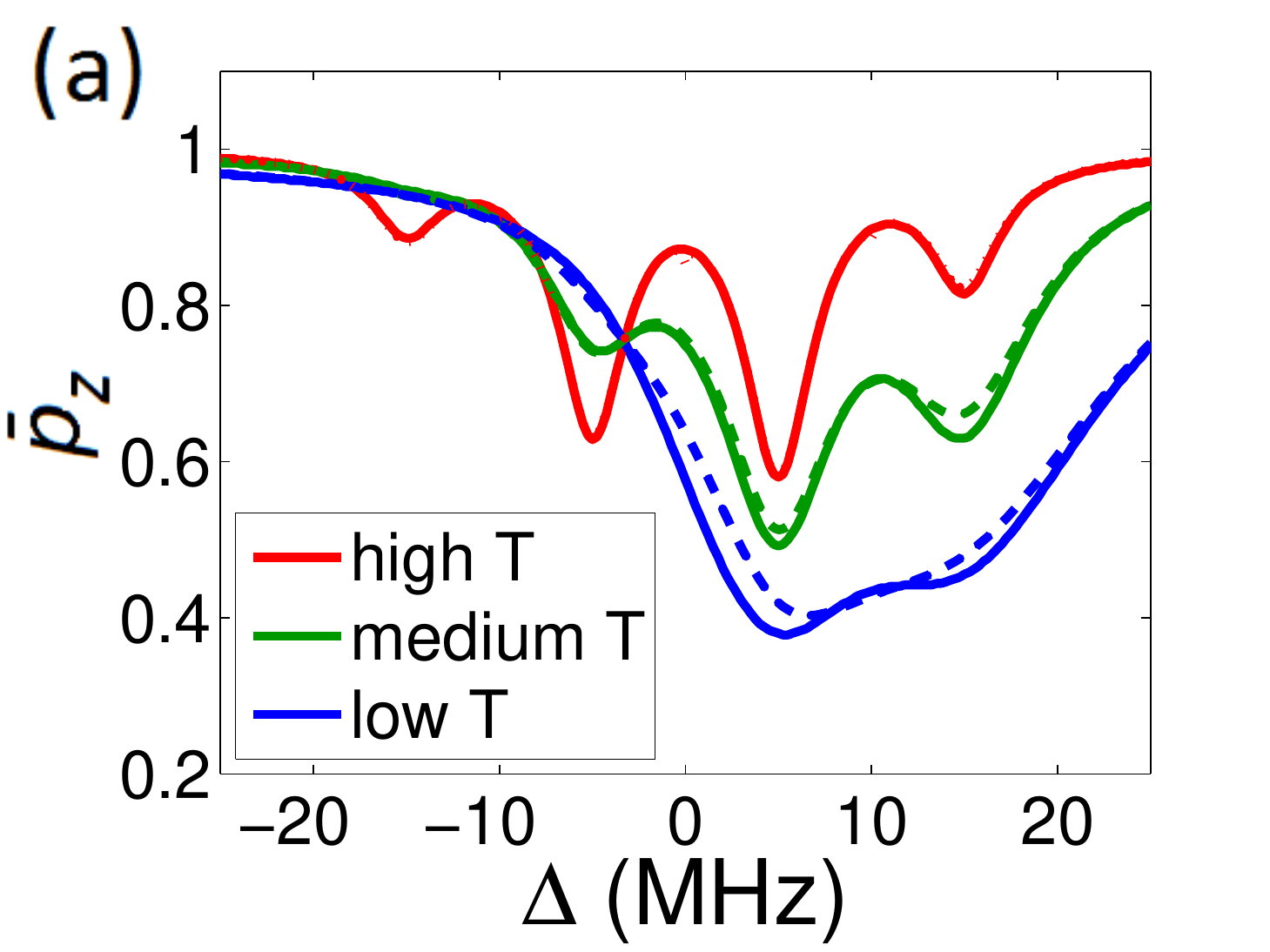}
\includegraphics[scale=.3]{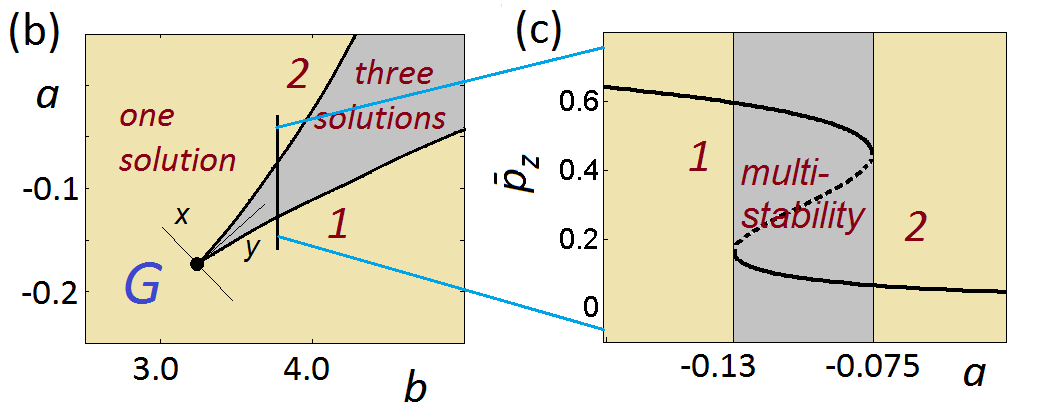}
\includegraphics[scale=.3]{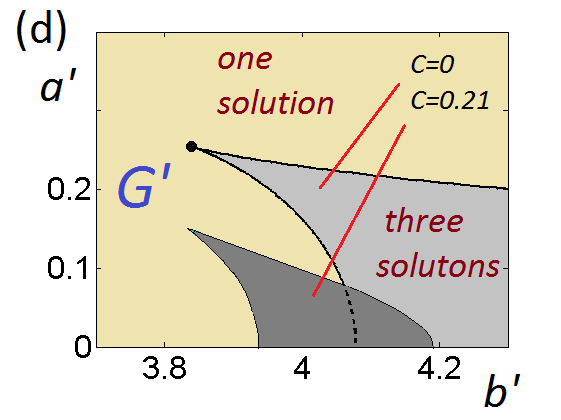}
\end{center}
\caption{\label{F}
(a) Steady-state polarization spectra $\bar p_z(\Delta)$ obtained by the mean-field formula (\ref{mf}) (solid lines) and the numerically exact solution (dashed lines) for $N=4$, $D=10$ MHz, $R_2=10^6\ {\rm s}^{-1}$ and different temperature and microwave parameters: $p=0.11$, $\omega_1=75$ kHz, $R_1=10^3\ {\rm s}^{-1}$ (red); $p=0.55$, $\omega_1=12$ kHz, $R_1=10\ {\rm s}^{-1}$ (green); $p=0.99$, $\omega_1=7$ kHz, $R_1=1\ {\rm s}^{-1}$ (blue). (b) Phase diagram obtained from Eq.~(\ref{mf}) in the $(a,b)$-plane. The diagram features regions of unique (brown) and multiple (gray) solutions and displays a (cusp) critical point $G$ at $p=0.99$, $\omega_1=R_2=10^5$ and $R_1=1\ {\rm s}^{-1}$ (for $N=150$ electrons). (c) Structure of the solutions along the cut $b=3.75$ ($D=6.3$ MHz) through the region with multi-stable region featuring three solutions. (d) Phase diagram obtained from Eq.~(\ref{mfg}) in the $(a',b')$-plane featuring regions of unique and multiple solutions similar to that in panel (b) and a critical point $G'$ belonging to the same universality class as $G$ (see text for details). The dark gray region illustrates the contraction of the multi-stability region caused by inhomogeneous broadening (see text and Appendix \ref{CPD} for details).}
\end{figure*}

Dissipative processes within the electron system are modeled by the dissipator $\mathcal{D}$ which describes single-spin relaxation and takes the form
\begin{equation}
\begin{array}{c}
{\displaystyle\mathcal{D}=\sum_k\left[\gamma_{1+}{\cal L}(S_{k+})+\gamma_{1-}{\cal L}(S_{k-})+\gamma_2{\cal L}(S_{kz})\right],}\\
{\displaystyle\gamma_{1\pm}=\frac{R_1}{2}(1\mp p),\quad
\gamma_2=2R_2,\quad
p=\tanh\frac{\hbar\omega_S}{2k_\mathrm{B}T}}
\end{array}
\label{G}
\end{equation}
where ${\cal L}(X)\rho\equiv X\rho X^\dagger-\left\{X^\dagger X,\rho\right\}/2$ is the Lindblad form of a dissipation operator \cite{k-15}. The dissipation rates depend on the longitudinal ($R_1$) and transversal ($R_2$) relaxation rates of the electron spins as well as the thermal polarization $p\in[0,1]$. Here, $p$ is a function of the average electron Larmor frequency $\omega_S$ and the temperature $T$. For typical experimental conditions ($W$-band, $\omega_S\sim 100~{\rm GHz}$, sample temperature between $T\sim 0~{\rm K}$ and $T\sim 100~{\rm K}$) the thermal spin polarization takes on values between $p\sim 1$ and $p\sim 10^{-2}$.

\noindent
{\bf \em Mean-field in the absence of disorder} --- In order to  obtain a basic understanding of the phase structure of the driven electron system, let us first disregard any dispersion in the frequency offsets and interactions, by setting $\Delta_k = \Delta$ and $D_{kk'} = D/(N-1)$. (Note, that this $N$-dependence takes into account the fact that in practice the coupling strengths decay rapidly with the interspin distance and keep the interaction energy an extensive quantity.) In the non-disordered case, the last term of Eq. (\ref{h}) commutes with the rest of the Hamiltonian and does not influence the bulk polarization dynamics. Therefore, we can neglect it, leading to the mean-field Hamiltonian
\begin{equation}
\bar H=\sum_k\left(\omega_1S_{kx}+\Delta S_{kz}\right)+\frac{3D}{N-1}\sum_{k<k'}S_{kz}S_{k'z}.
\label{H'}
\end{equation}

We now compute the stationary average bulk polarization $p_z=-2\sum{\rm Tr}\,(S_{kz}\rho_{\rm ss})/N$ which serves as an order parameter for classifying the steady state $\rho_{\rm ss}$. To obtain the mean-field equation we define $H_k=\omega_1S_{kx}+\bar\Delta_kS_{kz}$ which is the projection of $\bar H$ onto the subspace of spin $k$. Here $\bar\Delta_k=\Delta+\frac{3D}{N-1}\sum_{k'\not=k}S_{k'z}$ is the effective energy shift or offset term experienced by the spin. This takes discrete values, i.e.,
\begin{equation}
  \bar\Delta_k \,\, \in\,\,\delta(q)=\Delta+\frac{3D}{N-1}\left(q-\frac{N-1}{2}\right)
\end{equation}
where $q=0,...,N-1$ is the number of spins $k' \neq k$ in the up-state. The steady-state polarization $p'_z(q)$ of a single spin for given $q$ is [see Appendix \ref{SSSP}]
\begin{equation}
p'_z(q)=p\,\left(1-\frac{\eta\omega_1^2}{\delta_0^2+\delta^2(q)}\right)
\label{p'}
\end{equation}
where $\delta_0=\sqrt{R_2^2+\eta\omega_1^2}$ and $\eta=R_2/R_1$ is the ratio of the electron spin relaxation rates. Since the system is homogeneous, the steady-state polarization of the individual spins is identical and given by $p_z$, which can be regarded as a self-consistency condition. Hence, the probability of having $q$ up spins and $N-q-1$ down spins is given by $P(q,p_z)=\binom{N-1}{q}\frac{(1-p_z)^q(1+p_z)^{N-1-q}}{2^{N-1}}$. Averaging Eq.~(\ref{p'}) over all values of $q$ finally yields the equation for the relative steady-state polarization $\bar p_z=p_z/p$:
\begin{equation}
\bar p_z=f(\bar p_z)\equiv\sum_{q=0}^{N-1}P(q,p \, \bar p_z)p'_z(q)/p.
\label{mf}
\end{equation}

\noindent
{\bf \em Low and high temperature regime} --- The relative polarization is bounded ($\vert\bar p_z\vert\le 1$), thus $f(\bar p_z)$ defines a continuous map of the unit interval $\bar p_z\in[0,1]$ to itself. Therefore, by virtue of the Brouwer fixed point theorem \cite{b}, Eq.~(\ref{mf}) always has at least one solution. We find that the solution is unique for small values of $p$ corresponding to high temperatures and small numbers of spins $N$ (see Appendix \ref{Uniqueness}).

For small values of $N$ we can compare the results of the mean-field treatment to the exact solution of the quantum master equation given by the dissipator (\ref{G}) and Hamiltonian (\ref{H'}). To this end we show in FIG.~\ref{F}\textcolor{blue}{(a)} the steady-state polarization spectrum, i.e. the dependence of the bulk polarization $\bar p_z$ on the average microwave offset $\Delta$, for three typical sets of parameters for $N=4$. Generally a good agreement is obtained. The observed spectra have $N$ Lorentzian peaks occurring at $\Delta=3D(1/2-q/(N-1))$, $q=0,1,\ldots,N-1$, with a half-width of $\delta_0$. The centre $\Delta=0$ of the spectrum corresponds to $q\sim q_0\equiv(N-1)/2$. The mean of the binomial distribution $P(q,p \, \bar p_z)$ where the maximal saturation is given by $\bar q=(N-1)(1-p \, \bar p_z)/2$. Here $\bar q$ is close to $q_0$ for small $p$ and tends to shift from $q_0$ with increasing $p$. Hence, the intensities of the peaks are symmetric with respect to the centre of the spectrum at high temperatures ($p\sim 0$) and undergo a shift from the centre at low temperatures ($p\sim 1$).

For large $N$ and high temperature we find a single broad region around $\Delta\sim 0$, in which the polarization is saturated due to the applied field. The width of this region increases with interaction strength $D$ (see Appendix \ref{Uniqueness} for details).

\noindent
{\bf \em Multi-stability and phase transitions} --- The situation qualitatively changes when entering the regime of low temperatures $p\sim 1$ and high numbers of spins $N\gg 1$. In this case (see Appendix \ref{Uniqueness}) Eq.~(\ref{mf}) can feature more than one solution. In FIG.~\ref{F}\textcolor{blue}{(b)} we show the phase diagram given by the number of solutions of Eq.~(\ref{mf}) in terms of the scaled offset and interaction parameters $a=\Delta/\omega_1\sqrt{\eta}$, $b=3D/\omega_1\sqrt{\eta}$. FIG. \ref{F}\textcolor{blue}{(b)} features a multi-stability region where three solutions coexist (gray) separated from the regions with a unique solution (brown) by two spinodal lines that coalesce at a critical point $G$. Similar phase diagrams have recently been discussed theoretically in other contexts, e.g., for open driven gases of strongly interacting Rydberg atoms \cite{m,Carr2013,deMelo2016,Weller2016}, or certain classes of dissipative Ising models \cite{Ates2012Ising,Weimer2015,Rose2016}. The behavior of the steady-state polarization $\bar p_z$ upon crossing the multi-stable region is shown in FIG.~\ref{F}\textcolor{blue}{(c)}.

Solutions with small $\bar p_z\sim 0$ correspond to non-thermal quasi-saturated equilibrium states. States with large values $\bar p_z\sim 1$ are unsaturated quasi-thermal equilibria. On crossing the spinodal curve $1$ from large negative values of $a$, the unique stable quasi-thermal steadystate continues to exist but two other steadystate solutions appear: a stable quasi-saturated and an unstable intermediate one as shown in FIG. \ref{F}\textcolor{blue}{(c)}. Conversely, on crossing curve $2$ towards large negative values of $a$, the unique stable quasi-saturated steadystate continues to exist but two other steadystates emerge, a stable and an unstable one.

\begin{figure*}[t]
\includegraphics[scale=.54   %width=1\linewidth
]{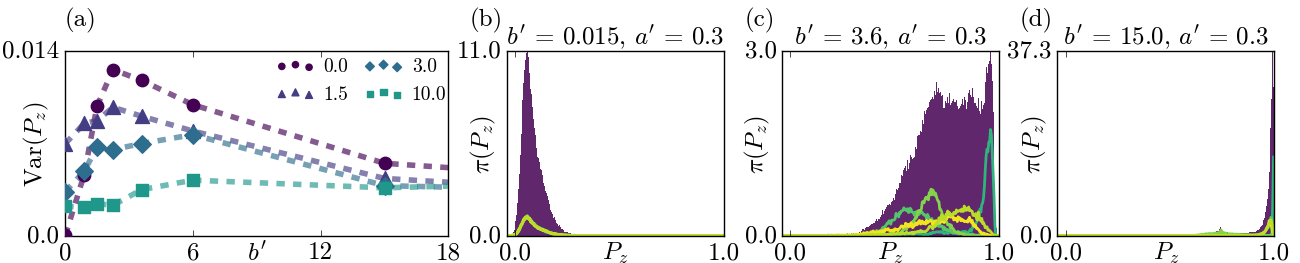}
\caption{Numerical simulations and fluctuations. All results in this figure are produced for parameters $\omega_1=10^5\text{Hz}$, $R_1=1\text{s}^{-1}$, $R_2=10^5\text{s}^{-1}$, $p=0.99$ and $N=8$, and averaged over 10 disorder realizations. (a) The variance of the time integrated observable ${P}_{z}$ for varying $b'$, with the fixed $a'$ value indicated by the legend in the top right. (b-d) Discrete approximations of the probability density (dark shaded area) for the observable $P_z$ for three sets of parameters, such that $\int\pi(P_z)dP_z=1$ over the range shown. The light colored curves represent the densities for some individual disorders, divided by the number of disorder realizations considered so that their addition (rather than their average) would equal the full probability density. This is done to better represent the contribution each disorder realization makes to the distribution.}\label{QJMCResults}
\end{figure*}

The occurrence of multiple steady state solutions is an artifact of the mean-field approximation.  It can be interpreted as the emergence of metastable states \cite{Rose2016} near first-order phase transitions.  An experimental signature of this type of physics is for example hysteretic behavior as recently studied in the context of interacting atomic gases \cite{Carr2013,deMelo2016,Weller2016}. We will return to this point further below.

The nature of the critical point $G$ in the phase diagram FIG.~\ref{F}\textcolor{blue}{(b)} can be characterized by analysing the scaling behaviour of $\bar p_z$ near it. We find two directions that are singled out (see Appendix \ref{CPD} for details): one is given by the curve that is tangent to both spinodal lines [see FIG.~\ref{F}\textcolor{blue}{(b)}], where we find $|{\bar p_z}-{\bar p}_\mathrm{crit}| \sim y^{1/2}$, where ${\bar p}_\mathrm{crit}$ is the value of ${\bar p_z}$ at the critical point. Along the perpendicular direction we find $|{\bar p_z}-{\bar p}_\mathrm{crit}| \sim x^{1/3}$. These are  Ising mean-field exponents \cite{goldenfeld}. In the context of a classical Ising model, the directions $x$ and $y$ would correspond to magnetic field and temperature respectively (see also Ref. \cite{m}).

\noindent
{\bf \em Disordered spin-spin interactions and augmented mean-field} --- The results so far indicate possible phase transitions in the polarization of the electron system  controlled by the frequency offset $\Delta$ and the average interaction strength $D$. However, typical sample materials are not single crystals and electrons are arranged randomly, such that the average interaction experienced by an electron is close to zero \cite{k-16}. In order to take this into account we need an augmented mean-field description which accounts for a distribution in the coupling strengths.

Note that when the disorder in either the offsets $\Delta_k$ or the interactions $D_{kk'}$ is large enough, unitary dynamics with Hamiltonian (\ref{h}) is expected to undergo many-body localisation (MBL) \cite{Nandkishore2015}.  In this case spatial fluctuations in the long-time state can be significant and determined by the disorder and the initial state, which raises the question of the appropriateness of mean-field.  However, in the presence of dissipation, cf.\ Eq.~(\ref{G}), MBL is unstable and the stationary state is delocalised  \cite{Levi2016,Medvedyeva2016,Fischer2016}, suggesting that the mean-field analysis is still appropriate.  (For other possible connections between MBL and DNP see \cite{lr-15}.)

For the sake of simplicity we assume that the interactions $D$ follow a Gaussian distribution, $\chi(D)=\exp(-D^2/D_0^2)/(\sqrt{\pi} D_0)$, with zero mean and standard deviation $D_0$.  The offset frequency $\Delta$ may also be disordered (e.g.,  from the $g$-anisotropy and hyperfine interactions with nuclei \cite{sj-01,pf-79}), but we neglect that effect for now.  Eq.~(\ref{mf}) generalizes to
\begin{equation}
\bar p_z=\int_{-\infty}^{+\infty}f_0(D,\bar p_z)\chi(D)\, dD.
\label{mfg}
\end{equation}
Here we replaced the function $f(\bar p_z)$ by its average with respect to the distribution $P(q,p \, \bar p_z)$: $f_0(D,{\bar p_z})=p'_z(\bar q)/p=1-\eta\omega_1^2/(\delta_0^2+\delta^2)$ with $\delta=\Delta-3Dp \, \bar p_z/2$. This is justified by the properties of the distribution $P$ and by the fact that the averaged function $f_0$ coincides with the classical mean-field approximation of the Ising model in the limit of $N\gg 1$, so Eq.~(\ref{mfg}) no longer depends on $N$ \cite{Rose2016,Ates2012Ising,m} (see also Appendix \ref{CMTI} for details). The mean-field phase diagram resulting from Eq.~(\ref{mfg}) is displayed in FIG.~\ref{F}\textcolor{blue}{(d)} as a function of the dimensionless parameters $a'=\Delta_0/\omega_1\sqrt{\eta}$ ($\Delta_0$ is the average offset, equal to $\Delta$ in the case considered here) and $b'=3pD_0/2\omega_1\sqrt{\eta}$. We assume that the strength of the microwave field is large: $\omega_1^2\eta\gg R_2^2$ meaning that the electron system is fully saturated in the absence of spin-spin coupling (in which case the phase transitions observed are most pronounced). The structure is similar to that of FIG.~\ref{F}\textcolor{blue}{(b)}. We observe regions with one and three solutions as well as spinodal lines forming a cusp at a critical point $G'$. The scaling properties at this critical point are, again, those of mean-field Ising universality. Note, however, that the phase transition is controlled by the width of the distribution of the disorder strengths ($D_0\propto b'$), rather than the average interaction strength, which is in fact zero.

Eq. (\ref{mfg}) can be modified to take into account disorder in the frequency offsets $\Delta_k$. To this end the probability density $\chi(D)$ in Eq. (\ref{mfg}) is replaced by a joint probability density $\chi(D,\Delta)$ accounting for both homogeneous and inhomogeneous broadening. The disorder in $\Delta_k$ causes a shift and contraction of the multi-stability region which is illustrated by the dark gray region in FIG.\ref{F}\textcolor{blue}{(d)} where the dimensionless parameter $c$ characterizes inhomogeneous broadening (see Appendix \ref{IB} for details).

\noindent
{\bf \em Fluctuations and numerical simulations} --- The mean-field treatment above is of course not exact.  Whether the predicted qualitative phase structure survives away from mean-field depends on the effect of fluctuations \cite{SchirmerWang2010,Weimer2015}. As shown in \cite{Rose2016,Ates2012Ising,foss2016emergent}, phase coexistence at the mean-field level can be an indication -- away from the thermodynamic limit -- of the existence of long-lived metastable (rather than stationary) phases. These competing phases come with an intermittent dynamics of slow switching between them.  We now show that this is indeed the case by considering the dynamics of the exact system, Eqs.~(\ref{h}), (\ref{G}), by means of numerical simulations in small systems.

We study the time dependence of the polarization ${p}_{z}(t)=-(2/N)\sum_k{\rm Tr}\,(S_{kz}\rho(t))$ for a variety of values of $a'$ and $b'$. For the set of parameters we consider, multiple disorder realizations of the dipolar coupling $\{D_{kk'}\}$, with $D'_{kk'}=D_{kk'}$ are taken. These are independent and identically distributed, sampled from a Gaussian distribution with variance defined by $b'$ (see Appendix \ref{QJMC} for details). Fluctuations are quantified through the variance of the integrated polarization, $P_z=1/t \int_{0}^t p_z(t')dt'$. In our simulations $t$ is chosen long enough, such that fluctuations due to the transient, short time dynamics average out. In FIG.~\ref{QJMCResults}\textcolor{blue}{(a)} we show the disorder averaged variance of $P_z$ as a function of $b'$ for several values of $a'$, cf.\ FIG.~\ref{F}\textcolor{blue}{(d)}. All curves display a peak indicating enhanced fluctuations for intermediate values of $b'$, which is the region where metastable states and enhanced fluctuations are expected. Similar behaviour is observed in FIG.~\ref{QJMCResults}\textcolor{blue}{(b-d)} for the probability distribution of $P_z$, shown both for individual disorders and averaged over disorder. Since the system is small, we do not expect self-averaging, and $P_z$ for individual realisations of the disorder to vary.  Nevertheless, all histograms broaden significantly for intermediate values of $b'$, clearly displaying enhanced fluctuations as indicated by the multi-stable region identified by our mean-field analysis.

\noindent
{\bf \em Conclusions} --- Our results demonstrate that cooperative behaviour
in strongly interacting ensembles of microwave driven electrons - a situation of relevance to DNP in NMR - can give rise to a non-trivial phase structure in the stationary state of these systems. Mean-field analysis predicts the existence of phases of distinct polarisation, with phase transitions between them controlled by the detunings in the microwave driving and the distribution of the dipolar electronic couplings. While the calculated phase diagram is mean-field in origin, our simulations show that -- even for finite systems -- dynamics will be correlated and intermittent, related to the coexistence of metastable states. The experimental demonstration of these predicted phenomena would ideally require a paramagnetic sample with minimal inhomogeneous broadening, kept at cryogenic temperatures and high magnetic field.

\begin{acknowledgments}
%{\it Acknowledgments} --- 
The authors thank B. Olmos and J. A. Needham for useful discussions. The research leading to these results has received funding from the European Research Council under the European Union's Seventh Framework Programme (FP/2007-2013) / ERC Grant Agreement No. 335266 (ESCQUMA) and the EPSRC Grant No. EP/N03404X/1. We are also grateful for access to the University of Nottingham High Performance Computing Facility.
\end{acknowledgments}

\appendix

\section{Steady-state of single-spin microwave-driven dynamics}\label{SSSP}
In the context of our work, the microwave-driven single-spin master equation has the form
$$
\dot\rho=-i[H,\rho]+\mathcal{D}\rho
$$
with
$$
H=\omega_1S_x+\delta S_z,
$$
$$
\mathcal{D}=\frac{R_1}{2}\left[(1-p){\mathcal L}(S_{+})+(1+p){\mathcal L}(S_{-})\right]+2R_2{\mathcal L}(S_{z}).
$$
In terms of the relative polarization components
$$
\rho=1/2-p\left(XS_x+YS_y+ZS_z\right),
$$
we come to the Bloch equations (for $R_2\gg R_1$) 
$$
\dot X=-\delta Y-R_2X,\quad
\dot Y=\delta X-\omega_1Z-R_2Y,
$$
$$
\dot Z=\omega_1Y+R_1(1-Z).
$$
The steady-state solution where the right-hand sides are all zero is unique and calculated as
$$
X=\frac{\omega_1\delta}{R_2^2+\delta^2}\,Z,\quad
Y=-\frac{\omega_1R_2}{R_2^2+\delta^2}\,Z,
$$
$$
Z=1-\frac{\omega_1^2\eta}{\delta_0^2+\delta^2},\quad
\delta_0^2=R_2^2+\omega_1^2\eta,\quad
\eta=\frac{R_2}{R_1}
$$
in full agreement with Eq. (\ref{p'}). 

\section{Uniqueness of solution for \\high temperatures and small \\numbers of spins}\label{Uniqueness}

\begin{figure}[t]
	\begin{center}
		\includegraphics[scale=0.28]{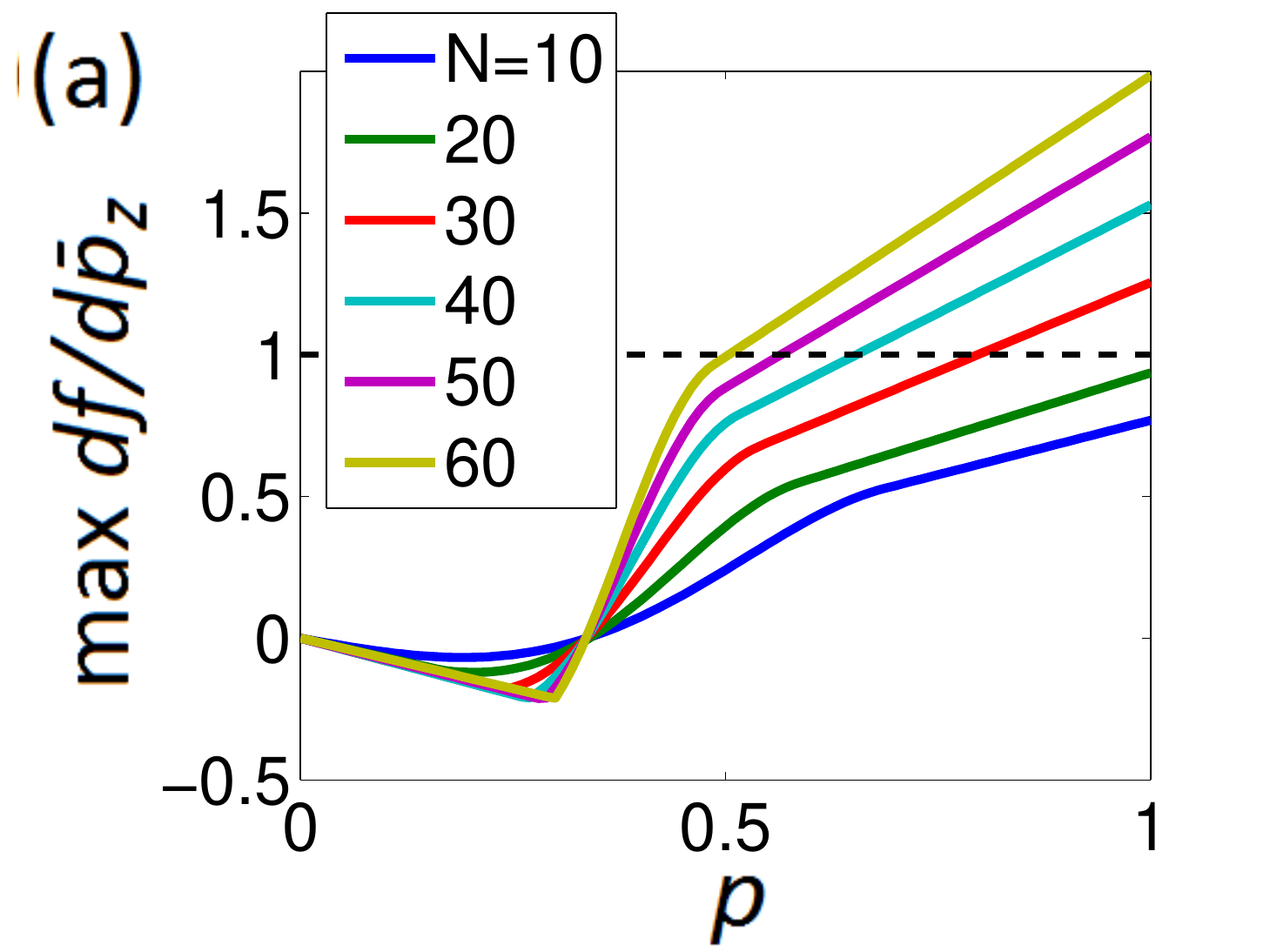}
		\includegraphics[scale=0.28]{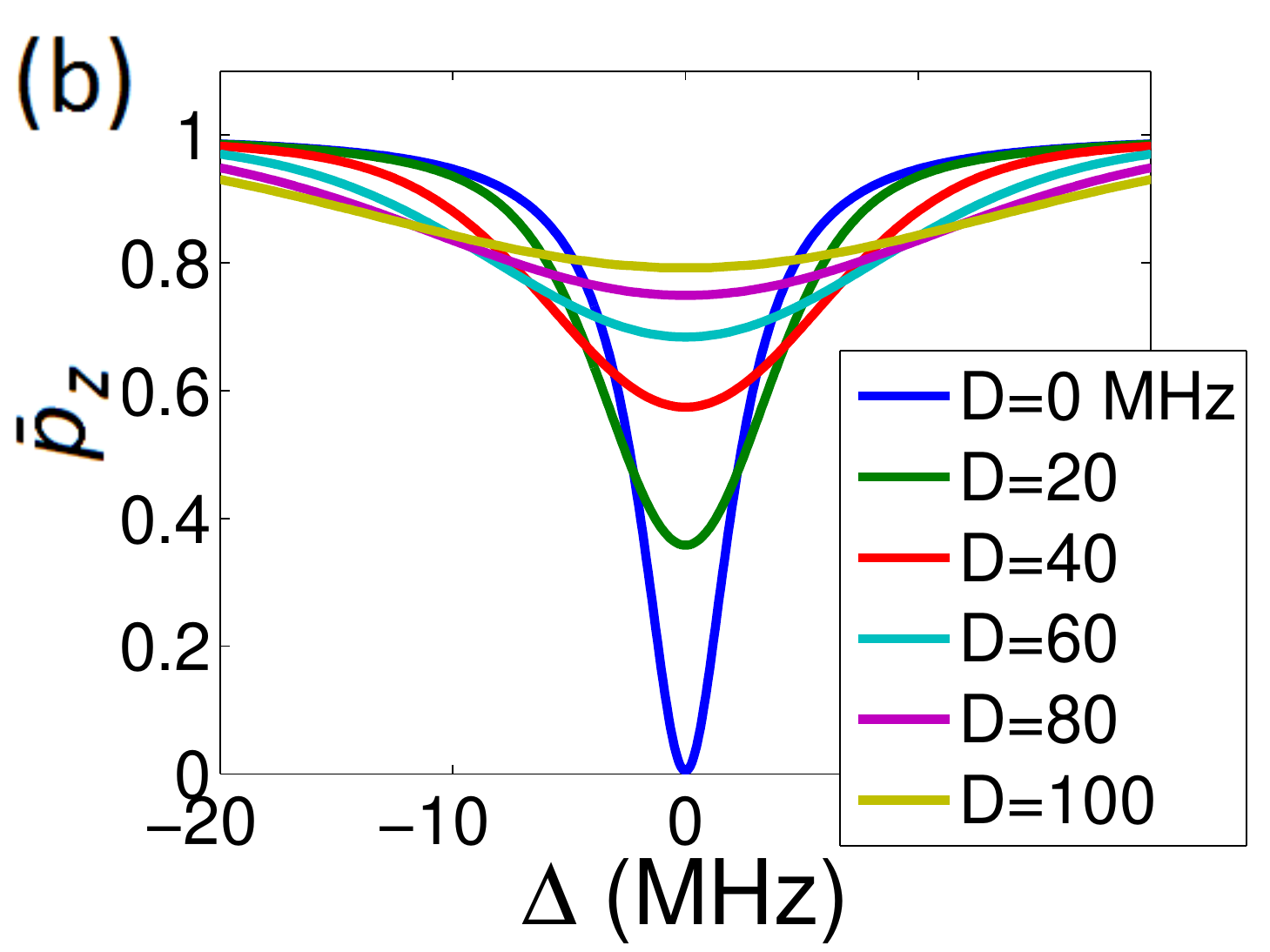}
	\end{center}
	\caption{\label{SM1}
		(a) Dependence of $\max df/d\bar p_z$ on the thermal polarization $p$ for different values of $N$ at $\Delta=10$ MHz, $D=20$ MHz.
		(b) High-temperature steady-state polarization spectra for different values of $D$, calculated with Eq. (\ref{mf}) for $N=150$. In both panels,  other parameters are chosen as in the red curve of FIG.\ref{F}\textcolor{blue}{(a)} of the main text.}
\end{figure}

To understand the structure of the solution space of Eq. (\ref{mf}) as a function of the thermal polarisation $p$ and the number of electrons $N$, we consider the derivative $df/d\bar p_z$: it is proportional to $p$, and thus for small values of $p$, corresponding to high temperatures, we have $df/d\bar p_z<1$. Under this condition the graph of the function $f(\bar p_z)$ can intersect the diagonal $g(\bar p_z)=\bar p_z$ only once and hence Eq. (\ref{mf}) has only one solution. This high temperatures behaviour is independent of the number of spins $N$, which is illustrated in FIG.~\ref{SM1}\textcolor{blue}{(a)}. Here we plot $\max_{\bar p_z}df/d\bar p_z$ as function of $p$ for different values of $N$ and fixed other parameters, showing that the maximum slope for small $p$ is always negative. The shape of the steady-state polarization spectrum $\bar p_z(\Delta)$ is described and good agreement between the master equation and the meanfield Eq. (\ref{mf}) for small $N$ is illustrated in the main text. In FIG.~\ref{SM1}\textcolor{blue}{(b)} we show the high-temperature steady-state polarization spectrum resulting from Eq. (\ref{mf}) for large $N$ and different values of $D$. Broadening of the saturation region around $\Delta=0$ with increasing $D$ is evident. 

\section{Structure of the phase diagram}\label{CPD}

Mathematically, the phase diagram of a (smooth) general two-parametric family of self-consistent relations of the form 
\begin{equation}
u=f(a,b,u),
\label{scr}
\end{equation}
can be studied from the point of view of the singularities in geometry of the 2-dimensional surface defined by the relation (\ref{scr}) in the 3-space $(a,b,u)$. The relation (\ref{scr}) can be rewritten as
$$
u-f(a,b,u)=\frac{\partial F}{\partial u}=0,\quad
F=\frac{u^2}{2}-\int f(a,b,u)\,du,
$$
which defines a critical point $u$ of a (smooth) scalar function $F(u)$ depending on the parameters $a,b$. This makes a subject of the mathematical theory of singularities combined with the geometry of the surface (\ref{scr}) known as the catastrophe theory \cite{a-84}.  

Consider the Taylor expansion of Eq. (\ref{scr}) near a given value $u=u^*$
$$
u^*+v=f(a,b,u^*+v)=f(a,b,u^*)+\frac{\partial f}{\partial u}(a,b,u^*)v+
$$
$$
\frac{1}{2}\,\frac{\partial^2f}{\partial u^2}(a,b,u^*)v^2+\frac{1}{6}\,\frac{\partial^3f}{\partial u^3}(a,b,u^*)v^3+\ldots\equiv
$$
$$
c_0+c_1v+c_2v^2+c_3v^3+\ldots
$$
If $c_0\not=u^*$ then near the value $u=u^*$ Eq. (\ref{scr}) does not have solutions. If $c_0=u^*$ then $u=u^*$ is a solution, and we have
$$
v=c_1v+c_2v^2+c_3v^3+\ldots
$$
If $c_1\not=1$ then the solution $u=u^*$ is locally unique. If $c_1=1$, $c_2\not=0$ then $u=u^*$ is a degeneracy point where two solutions merge,
$$
0=c_2v^2+c_3v^3+\ldots
$$  
If $c_2=0$, $c_3\not=0$ then $u=u^*$ is a degeneracy point where three solutions merge, 
$$
0=c_3v^3+\ldots,
$$
etc. Since relation (\ref{scr}) depends on two parameters $a,\,b$ and one variable $u$, in a generic situation no more than three conditions on the coefficients $c_0,\,c_1,\,c_2$ can be simultaneously satisfied, so not more than three solutions can merge at $u=u^*$. The latter takes place at the so-called cusp point $G$ of the phase diagram \cite{a-84} which is defined by the critical values $a=a^*$, $b=b^*$, $u=u^*$ with
\begin{equation}
c_0=u^*,\quad
c_1=1,\quad
c_2=0,\quad
c_3\not=0
\label{cp}
\end{equation}
which means
$$
f(a^*,b^*,u^*)=u^*,\quad
\frac{\partial f}{\partial u}(a^*,b^*,u^*)=1,
$$
$$
\frac{\partial^2f}{\partial u^2}(a^*,b^*,u^*)=0,\quad
\frac{\partial^3f}{\partial u^3}(a^*,b^*,u^*)\not=0.
$$

\begin{figure}[t]
	\begin{center}
		\includegraphics[scale=0.5]{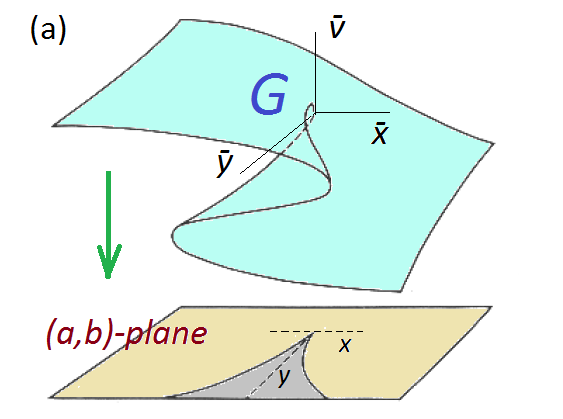}
		\vskip 5mm
		\includegraphics[scale=0.31]{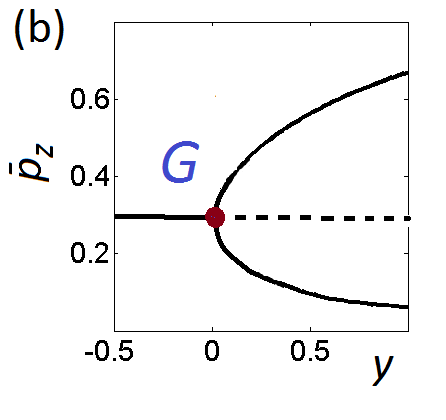}
		\includegraphics[scale=0.32]{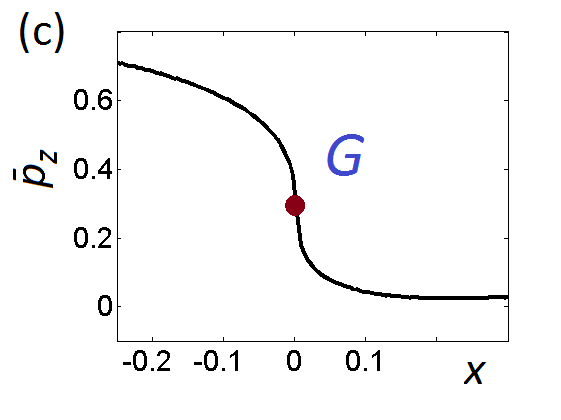}
	\end{center}
	\caption{\label{SM2}
		(a) Universal two-parametric phase diagram considered from the point of view of the mathematical catastrophe theory.
		(b) Structure of the solutions $\bar p_z$ of Eq. (\ref{mf}) on crossing the critical point $G$ along the tangent direction $y$. Two stable solutions separated by $\sim y^{1/2}$ are forked from the intermediate solution that loses its stability. 
		(c) The shape of the solution $\bar p_z$ on crossing the critical point along the perpendicular direction $x$, with a singularity $\sim x^{1/3}$.}
\end{figure}

Consider now the Taylor expansion of Eq. (\ref{scr}) near the cusp point up to terms of the third order, taking into account Eq. (\ref{cp}), 
$$
u^*+v=f(a^*+\alpha,b^*+\beta,u^*+v)\sim
$$
$$
u^*+\xi_0+(1+\xi_1)v+\xi_2v^2+\xi_3v^3+\ldots
$$
which implies
\begin{equation}
0\sim \xi_0+\xi_1v+\xi_2v^2+\xi_3v^3
\label{ce}
\end{equation}
with
$$
\xi_0=\frac{\partial f}{\partial a}\alpha+\frac{\partial f}{\partial b}\beta+
\frac{1}{2}\frac{\partial^2f}{\partial a^2}\alpha^2+\frac{\partial^2f}{\partial a\partial b}\alpha\beta+
\frac{1}{2}\frac{\partial^2f}{\partial b^2}\beta^2+
$$
$$
\frac{1}{6}\frac{\partial^3f}{\partial a^3}\alpha^3+\frac{1}{2}\frac{\partial^3f}{\partial a^2\partial b}\alpha^2\beta+
\frac{1}{2}\frac{\partial^3f}{\partial a\partial b^2}\alpha\beta^2+\frac{1}{6}\frac{\partial^3f}{\partial b^3}\beta^3,
$$
$$
\xi_1=\frac{\partial^2f}{\partial u\partial a}\alpha+\frac{\partial^2f}{\partial u\partial b}\beta+
\frac{1}{2}\frac{\partial^3f}{\partial u\partial a^2}\alpha^2+
$$
$$
\frac{\partial^3f}{\partial u\partial a\partial b}\alpha\beta+\frac{1}{2}\frac{\partial^3f}{\partial u\partial b^2}\beta^2,\quad
\xi_3=\frac{1}{6}\frac{\partial^3f}{\partial u^3},
$$
$$
\xi_2=\frac{1}{2}\left(\frac{\partial^3f}{\partial u^2\partial a}\alpha+\frac{\partial^3f}{\partial u^2\partial b}\beta\right),
$$
where the derivatives of $f$ are taken at $u=u^*$, $a=a^*$, $b=b^*$. The asymptotic cubic equation (\ref{ce}) has three solutions if  $\bar D<0$ and has one solution if $\bar D>0$, where the discriminant $\bar D$ is given by the expression
$$
\bar D=\frac{1}{27^2\xi_3^6}\left[\left(3\xi_1\xi_3-\xi_2^2\right)^3+\right.
$$
$$
\left.\frac{1}{4}\left(2\xi_2^3-9\xi_1\xi_2\xi_3+27\xi_0\xi_3^2\right)^2\right]=
\bar D_2+\bar D_3+\ldots
$$
where $\bar D_n$ is the term of the $n$th order in $\alpha$, $\beta$. The lowest order term is the quadratic term originated from $\xi_0^2$. This term forms the full square
$$
\bar D_2=\frac{1}{4\xi_3^2}(r\alpha+t\beta)^2,\quad
r=\frac{\partial f}{\partial a},\quad
t=\frac{\partial f}{\partial b}.
$$ 
Making the rotation on the $(\alpha,\beta)$-plane
$$
x=\frac{r\alpha+t\beta}{\sqrt{r^2+t^2}},\quad
y=-\frac{r\beta-t\alpha}{\sqrt{r^2+t^2}}
$$
and rewriting the cubic term $\bar D_3$ in the new parameters $x,\,y$, we obtain up to the third order 
$$
\bar D\sim s_0x^2-s_1y^3+s_2y^2x-s_3yx^2+s_4x^3
$$
where the coefficients $s_{0-4}$ are expressed via the derivatives of the function $f(a,b,u)$ at the cusp point. We have $s_0=(r^2+t^2)/4\xi_3^2>0$, so we can write
$$
\bar D\sim s_0x^2\left(1-\frac{s_3}{s_0}y+\frac{s_4}{s_0}x\right)-s_1y^3+s_2y^2x\sim 
$$
$$
s_0x^2-s_1y^3+s_2y^2x.
$$   
In other words, the critical curve $\bar D=0$ is asymptotically represented by the equation
$$
s_0x^2-s_1y^3+s_2y^2x=0.
$$ 
The last term can be removed by a shift transformation $x\to x+O(y^2)$ and neglecting a term $\sim y^4$, so this curve is asymptotically written as 
$$
s_0x^2-s_1y^3=0:\quad
y=\left(\frac{s_0}{s_1}\right)^{1/3}x^{2/3}.
$$    
This equation defines a cusp curve on the $(x,y)$-plane with two branches tangent to the $y$-axis at the cusp point $G$, see FIG.\ref{SM2}\textcolor{blue}{(a)} where the local geometry of the singular surface (\ref{scrg}) is shown. In the rotated local coordinates, the cubic equation (\ref{ce}) representing the relation (\ref{scr}) takes the form
\begin{equation}
\bar v^3-\bar y\bar v-\bar x=0,\quad
\bar x=2s_0^{1/2}x,\quad
\bar y=3s_1^{1/3}y.
\label{scrg}
\end{equation}
Inside the cusp region $s_0x^2-s_1y^3<0$, Eq. (\ref{scrg}) has three solutions, outside the cusp region $s_0x^2-s_1y^3>0$ only one solution exists. On crossing the cusp point $G$ along the $y$-axis, the unique solution $\bar v=0$ forks into three solutions $\bar v=0$ and $\bar v=\pm\,\bar y^{1/2}$. On crossing $G$ along the $x$-axis, the unique solution has a singularity $\bar v=x^{1/3}$. The described asymptotics are universal, i.e., valid for any two-parametric relation (\ref{scr}) as soon as it has a critical point where relations (\ref{cp}) hold \cite{a-84}. 

The critical point $G$ of the phase diagram of Eq. (\ref{mf}) satisfying Eq. (\ref{cp}) was found numerically to be
$$
a_*\sim-0.18,\quad
b_*\sim 3.23,\quad
\bar p_{\rm crit}\sim 0.27
$$ 
with the characteristic directions in the $(a,b)$-plane
$$
x\sim 0.99(a-a_*)-0.14(b-b_*),
$$
$$
y\sim 0.99(b-b_*)+0.14(a-a_*).
$$
In FIG.~\ref{SM2}\textcolor{blue}{(b)}, the structure of the solution $\bar p_z$ is shown on crossing the critical point $G$ along the tangent direction $y$, in FIG.~\ref{SM2}\textcolor{blue}{(c)} --- the same on crossing along the perpendicular direction $x$. 

The critical point $G'$ of the phase diagram of Eq. (\ref{mfg}) corresponds to
$$
a'_*\sim 0.26,\quad
b'_*\sim 3.83,\quad
\bar p'_{\rm crit}\sim 0.20
$$ 
with the characteristic directions (not plotted)
$$
x'\sim 0.97(a'-a'_*)+0.25(b'-b'_*),
$$
$$
y'\sim 0.97(b'-b'_*)-0.25(a'-a'_*).
$$

\section{Link to the classical meanfield \\theory of the Ising model}\label{CMTI}

As  shown in the main text, the projection of the averaged Hamiltonian of Eq.~(\ref{H'}) to the subspace of a randomly chosen spin $k$ is written as
$$
H_k=\omega_1S_{kx}+\bar\Delta_kS_{kz}
$$
where
$$
\bar\Delta_k=\Delta+\frac{3D}{N-1}\sum_{k'\not=k}S_{k'z}.
$$
The classical meanfield theory consists in replacing each operator $S_{k'z}$ by its bulk steady-state observable (see, for example, \cite{Rose2016,Ates2012Ising,m})
$$
-\frac{p_z}{2}=\frac{1}{N}\sum_k{\rm Tr}\,(S_{kz}\rho).
$$
This leads to the single-spin Hamiltonian
$$
\tilde H=\omega_1S_x+\bar\Delta S_z,\quad
\bar\Delta=\Delta-3Dp\bar p_z/2. 
$$
Applying Eq.~(\ref{p'}) justified in Appendix \ref{SSSP}, we obtain for the relative steady-state polarization
\begin{equation}
\bar p_z=f_0(\bar p_z),\quad
f_0=1-\frac{\omega_1^2\eta}{\delta_0^2+\bar\Delta^2}. 
\label{cmf}
\end{equation}
Up to differences in notations, this is the classical self-consistent relation for the steady-state of the Ising model driven by a transversal field \cite{Rose2016,Ates2012Ising,m}. 

The same result is obtained if we replace in Eq. (\ref{mf}) the summation over all $q$ by a single mean value of the binomial distribution $P(q,p\bar p_z)$
$$
\bar q=(N-1)\,\frac{1-p\bar p_z}{2}. 
$$
Indeed,
$$
\delta(\bar q)=\Delta+\frac{3D}{N-1}\left(\bar q-\frac{N-1}{2}\right)=\bar\Delta,\quad
\frac{p'_z(\bar q)}{p}=f_0.
$$

To justify the proceeding from the whole set $q=0,\,1,\,\ldots,\,N-1$ to the mean $\bar q$, rescale the integer variable $q$ by a new variable $\epsilon$ by the rule
\begin{equation}
\epsilon_q=\frac{q}{N-1}
\label{resc}
\end{equation} 
where $\epsilon_q=0,\,1/(N-1),\,\ldots,\,1$ defines a uniform subdivision of the unit interval. The probability density of the variable $\epsilon_q$ is the same binomial distribution $P(q,p\bar p_z)$ and the detuning $\delta$ becomes a function of $\epsilon$, 
$$
\delta(q)=\Delta+3D\left(\epsilon_q-\frac{1}{2}\right)\equiv\delta'(\epsilon_q).
$$ 
Due to rescaling (\ref{resc}), the mean and the variance of the distribution $\epsilon_q$ are the mean and the variance of the distribution $P$ divided by $(N-1)$ and $(N-1)^2$ respectively, so we obtain
$$
\bar\epsilon=\sum_{q=0}^{N-1}\epsilon_qP(q,p\bar p_z)=\frac{(N-1)(1-p\bar p_z)}{2(N-1)}=\frac{1-p\bar p_z}{2},
$$ 
$$
\sigma_\epsilon^2=\sum_{q=0}^{N-1}\left(\epsilon_q-\bar\epsilon\right)^2P(q,p\bar p_z)=
$$
$$
\frac{(N-1)(1-p^2\bar p^2_z)}{4(N-1)^2}=\frac{1-p^2\bar p^2_z}{4(N-1)}.
$$
In the limit $N\gg 1$, the variance $\sigma^2_\epsilon$ becomes zero, so the distribution $\epsilon_q$ is reduced to a single mean value $\bar\epsilon$ taken with the probablity 1. The summation over $q$ can be replaced by an integration over the unit interval with the probablity density represented by the Dirac delta-function $\tilde\delta(\epsilon-\bar\epsilon)$,
$$
f(\bar p_z)=\sum_{q=0}^{N-1}P(q,p\bar p_z)p'_z(q)/p=
$$
$$
\int_0^1\tilde\delta(\epsilon-\bar\epsilon)\left(1-\frac{\omega_1^2}{\delta_0^2+\delta^{'2}(\epsilon)}\right)\,d\epsilon=
$$
$$
p'_z(\bar q)/p=f_0.
$$
This justifies the classical meanfield theory (\ref{cmf}) as a thermodynamic $N\gg 1$ limit of the meanfield theory developed in the main text. 
\newline\newline\newline
\section{Effect of inhomogeneous \\ broadening}\label{IB}

To estimate the effect of inhomogeneous broadening, we considered a system represented by two Gaussian spin packets of the same zero mean and standard deviation $D_0$ separated by a difference $2\Delta'$ between the detunings. Here the Gaussian density $\chi(D)$ in Eq. (\ref{mfg}) remains unchanged while the function $f_0(D,\bar p_z)$ is modified as
$$
f'_0(D,\bar p_z)=\frac{1}{2}\left(f_+(D,\bar p_z)+f_-(D,\bar p_z)\right),
$$
$$
f_\pm(D,\bar p_z)=1-\frac{\eta\omega_1^2}{\delta_0^2+\delta_\pm^2},\quad
\delta_\pm=\Delta\pm\Delta'-\frac{3Dp\bar p_z}{2}.
$$
The effect of $\Delta'\not=0$ can be estimated varying the dimensionless parameter $\displaystyle c=\frac{\Delta'}{\omega_1\sqrt{\eta}}$. For $c\not=0$, the phase diagram in the $(a',b')$-plane still features multi-stable regions but the latter are shifted and contracted with growing $c$. The contraction of the multistability region is explained by the fact that large differences between electron Larmor frequencies tend to quench the spin interactions and thus quench the multiplicity of the solution of the self-consistent relation Eq. (\ref{mfg}).     

\section{Quantum Jump Monte \\Carlo simulations}\label{QJMC}

The simulations for FIG. \ref{QJMCResults} were done using the Quantum Jump Monte Carlo algorithm \cite{Daley2014} to calculate the stochastic evolution (trajectory) of the pure state of the system over time. While all trajectories are initialized in the same state, the all up configuration, data from a trajectory is only considered after sufficient time has elapsed that there is no memory of the initial state (we can be certain such a time scale exists for this finite system due to the results of \cite{SchirmerWang2010}), i.e. after the relaxation time. The remainder of the trajectory is then cut up in to time periods $T$ of $\mathcal{O}({10}^{-2}\text{s})$, chosen such that short time fluctations are averaged out so that only long time fluctuations influence the variance of the time integrated observable (similar to the approach used in Sec. III E of \cite{Rose2016}).

Different disorder realizations are handled as follows: we begin by taking a set of random numbers from a Gaussian distribution of unit variance, defining the realization. For a given value of $b'$ we then rescale all of these numbers by the associated value of the standard deviation $D_0$. As it can be shown that the probability density satisfies $p_{1}(x)dx=p_{D_{0}}(D_{0}x)d(D_{0}x)$ where the subscript represents the variance of the Gaussian, this rescaling provides us with an equivalent set of numbers that were effectively drawn from a distribution with standard deviation $D_0$.

%\bibliography{ESR_paper}
%\bibliographystyle{apsrev4-1}

%merlin.mbs apsrev4-1.bst 2010-07-25 4.21a (PWD, AO, DPC) hacked
%Control: key (0)
%Control: author (72) initials jnrlst
%Control: editor formatted (1) identically to author
%Control: production of article title (-1) disabled
%Control: page (0) single
%Control: year (1) truncated
%Control: production of eprint (0) enabled
%
\end{document}